\documentclass[a4paper,fleqn,usenatbib]{mnras}

\usepackage{newtxtext,newtxmath}
\usepackage[T1]{fontenc}
\usepackage{ae,aecompl}

\usepackage{graphicx}	% Including figure files
\usepackage{amsmath}	% Advanced maths commands
\usepackage{amssymb}	% Extra maths symbols
\title[Planetary atmosphere loss]{Atmospheric mass loss of extrasolar planets orbiting magnetically active host stars}

\author[S. Lalitha et al.]{
Lalitha Sairam,$^{1}$\thanks{E-mail: lalitha.sairam@iiap.res.in}
J. H. M. M. Schmitt,$^{2}$
and Spandan Dash$^{3}$
\\
$^{1}$Indian Institute of Astrophysics, II Block, Koramangala, Bangalore 560 034, India\\
$^{2}$Hamburger Sternwarte, Gojenbergsweg 112, 21029 Hamburg\\
$^{3}$Indian Institute of Science, C.V Raman Avenue, Yeshwantpur, Bangalore 560 012, India
}

\date{Accepted XXX. Received YYY; in original form ZZZ}

\pubyear{2017}

\begin{document}
\label{firstpage}
\pagerange{\pageref{firstpage}--\pageref{lastpage}}
\maketitle

\begin{abstract}
Magnetic stellar activity of exoplanet hosts can lead to the production of large amounts of high-energy emission,
which irradiates extrasolar planets, located in the immediate vicinity of such stars.  
This radiation is absorbed in the planets' upper atmospheres, which consequently  heat up and evaporate,
possibly leading to an irradiation-induced mass-loss. We present a study of the
high-energy emission in the four magnetically active planet-bearing host stars Kepler-63, Kepler-210, WASP-19, and HAT-P-11, based on new \emph{XMM-Newton} observations.  
We find that the X-ray luminosities of these stars are rather high with orders of magnitude
above the level of the active Sun.  
The total XUV irradiation of these planets is expected to be stronger than that of well studied hot Jupiters. 
Using the estimated XUV luminosities as the energy input to the planetary 
atmospheres, we obtain upper limits for the total mass loss in these hot Jupiters.  

\end{abstract}


%%%%%%%%%%%%%%%%%%%%%%%%%%%%%%%%%%%%%%%%%%%%%%%%%%

%%%%%%%%%%%%%%%%% BODY OF PAPER %%%%%%%%%%%%%%%%%%

\section{Introduction}

The discovery of the first exoplanet 51~Peg~b,  
orbiting around a Sun-like star, by \cite{mayor1995jupiter}  has become
the starting point of a new research area in astrophysics, i.e.,
exoplanetary science. 
Many more discoveries quickly followed, and in 2002 the first
detection of an atmosphere around an exoplanet (HD 209458 b) was 
accomplished \citep{charbonneau2002detection}. 
Since then both	 the methods of detecting exoplanets and the ability to 
characterize their physical properties have made rapid progress. 
The lower mass limit of detected exoplanets has constantly decreased 
and the detection of low-mass planets, i.e., the super-Earths with masses in the 
range of 1-10 Earth masses, has now become possible.
Naturally, the goal is to ultimately detect an Earth-like planet within the 
stellar habitable zone, i.e., that zone
which allows the presence of liquid water on the planet's surface;
that goal has been recently bolstered by the discovery of seven 
Earth-like planets around 
Trappist-1, three of which fall in the habitable zone \citep{gillon2017seven}. 

For the habitability of a planet, its ability to hold an atmosphere of its own 
for a significantly long period of time is of paramount importance.
While it was originally thought that atmospheric losses in exoplanets occur 
predominantly through Jeans escape, the observations of atmospheric mass loss in 
HD~209458~b \citep{vidal2003extended} 
show that this process can grossly underestimate the actual mass loss. In a seminal paper,  
\cite{lammer2003atmospheric} hypothesize that exoplanets 
may lose their atmospheres through a different mechanism, 
which relies on a hydrodynamic mass loss model originally proposed by \cite{watson1981dynamics} and 
can give rise to orders of magnitude larger mass loss rates.  
In this model the high-energy radiation is absorbed in the upper layers of the 
planetary atmosphere, which consequently heat up to temperatures in excess of 
10,000~K, leading to a hydrodynamic outflow similar to the classic Parker 
model of the solar wind. The calculation of the atmospheric mass loss in this model context 
thus requires an assessment of the exoplanetary high-energy environment and in particular, the amount of 
X-ray, FUV and UV radiation it receives from its host star. 
Furthermore, several chemical processes like photochemical decomposition, charge exchange and sputtering are induced, all of which 
can contribute to the mass loss. \cite{lammer2003atmospheric} and \cite{sanz2010scenario} show that this mechanism can 
produce mass loss rates which are corroborated by observations of actual exoplanetary mass loss in hot Jupiters with large amounts of hydrogen and helium in their atmospheres.

With the here presented study of hot-Jupiter hosts, we provide a small sample of rather
active stars, where the effects of planetary evaporation are expected to be
very large.  With our exploratory X-ray observations we obtain
a better insight into the hottest parts of the outer stellar atmospheres 
of these planet hosts. Our analysis specifically allows us to estimate 
the coronal temperature and  the X-ray activity level  of these stars, 
which in turn allow us to calculate rough estimates of the mass loss of the orbiting exoplanets. In \S 2 we describe our target stars, in \S 3 we 
describe the \emph{XMM-Newton} observation, our data analysis and our results.  A discussion and our conclusions are presented in \S 4.

\section{The sample stars}

In this article, we present the results of new \emph{XMM-Newton} X-ray observations 
of  four hot-Jupiter hosts, i.e., the stars Kepler-63, Kepler-210, WASP-19, and HAT-P-11. 
All these target stars are in the \emph{Kepler} planetary candidate sample for host 
stars with excess photometric modulation ($\sim 1\%$), i.e., they are presumably very active with large
amounts of high-energy radiation produced; we list all the known
properties of the {\bf investigated host stars and their
exoplanets} in Tab.~\ref{tab:tab1}.  
In the following, we provide a
more detailed description of our sample stars.

\begin{figure*} 
\begin{center}
\includegraphics[width=0.24\textwidth,clip]{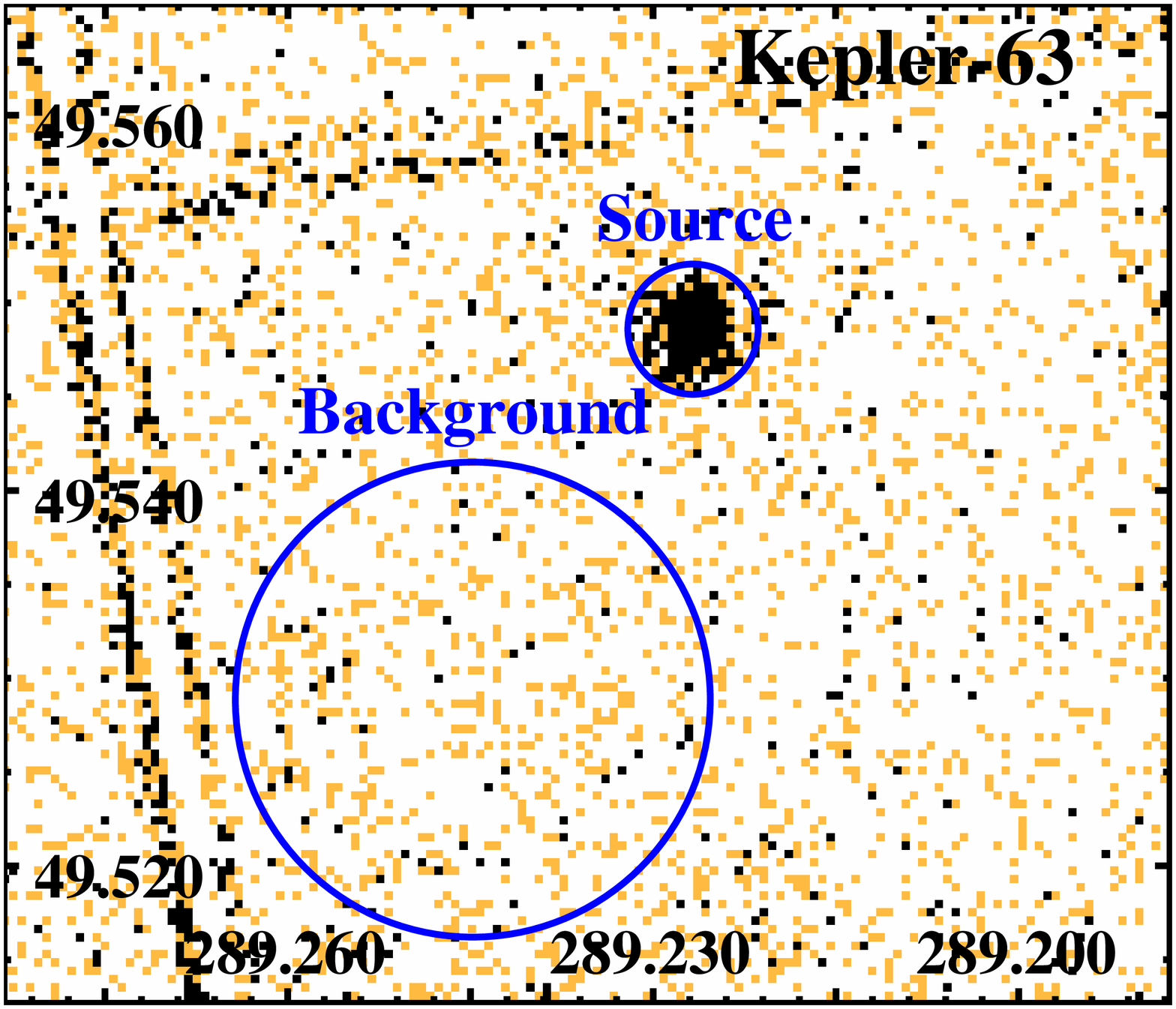} 
 \includegraphics[width=0.24\textwidth,clip]{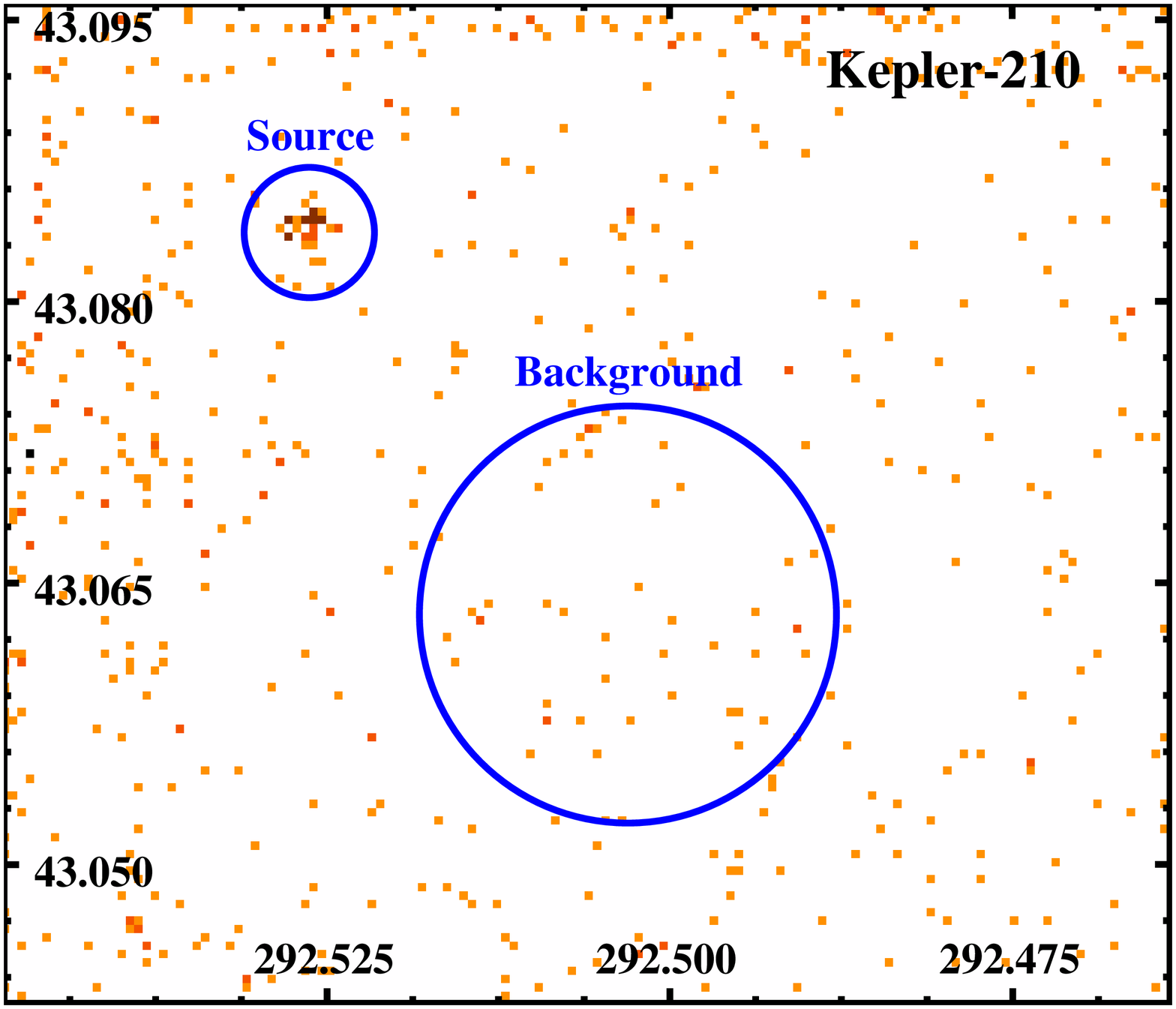}
\includegraphics[width=0.24\textwidth,clip]{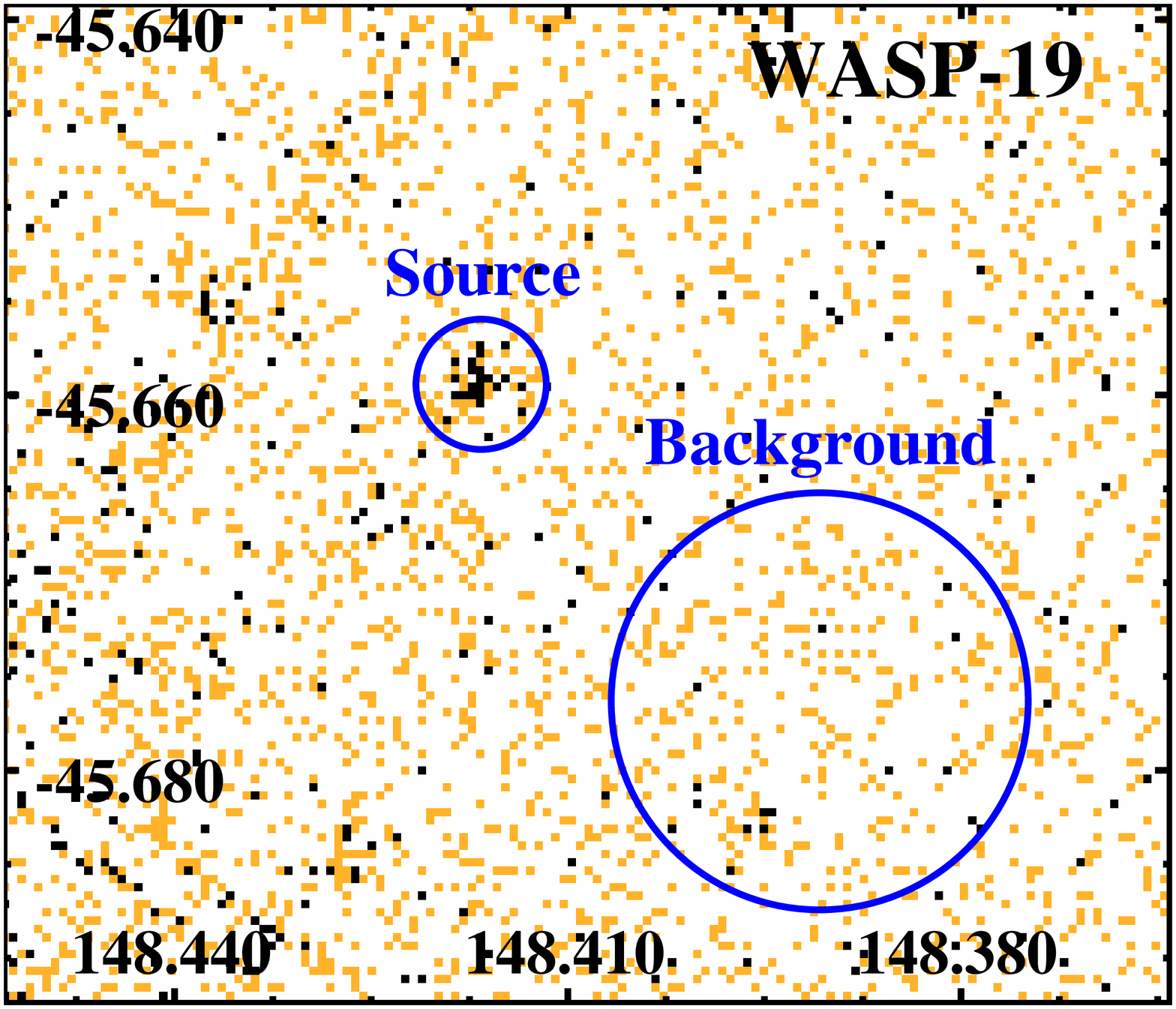}
\includegraphics[width=0.24\textwidth,clip]{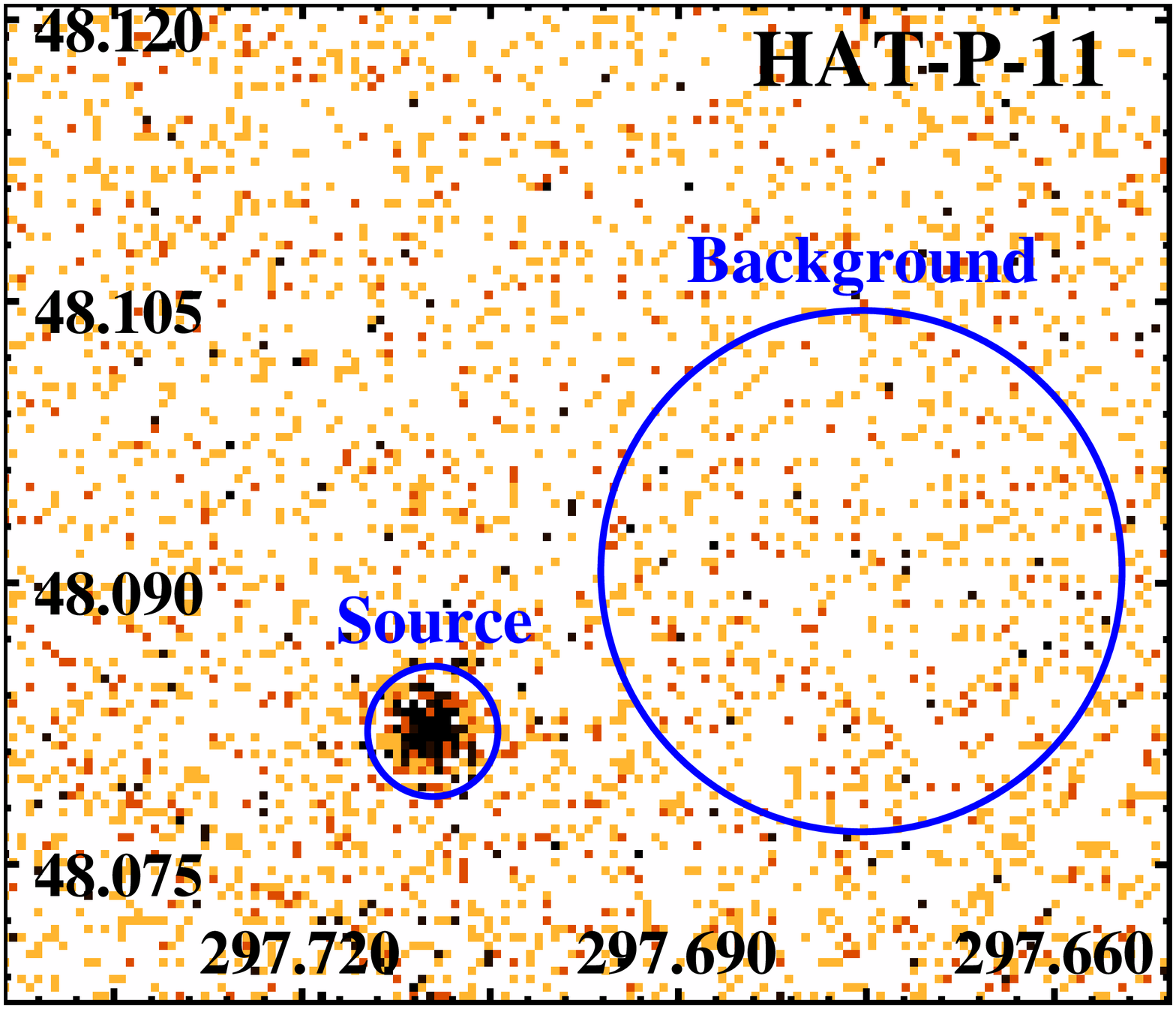}
\caption{\label{epicimg} Soft X-ray image of Kepler-63, Kepler-210, WASP-19 and HAT-P-11 from merged \emph{XMM-Newton} MOS1, MOS2 and PN camera in 0.2-2.0 keV energy band.}
\end{center}
\end{figure*}

\begin{table*}
\centering
\caption{Properties of star systems and the \emph{XMM-Newton} Observations}
\label{tab:tab1}
\begin{tabular}{lccccc} 
\hline\hline
Host stars & \textbf{Kepler-63} & \textbf{Kepler-210} & & \textbf{WASP-19}& \textbf{HAT-P-11}  \\ [0.5ex] 
\hline
\textbf{Distance (pc)} & $200 \pm 15$ & 265 &   & $250 \pm 80$ & $38 \pm 1.3$\\ 
\textbf{Spectral type} & G6 & M0V/K & & G8V & K4\\
\textbf{Temperature} & $5576\pm50$ K & $4559\pm100$ K & & $5500\pm100$ K &$4780 \pm 50$ K\\
\textbf{Age (Gyr)}&0.21$\pm$0.04&0.35$\pm$0.05&&0.6-5.5&6.5$^{5.9}_{4.1}$\\
\hline
Extrasolar Planets & Kepler-63b & Kepler-210c &  Kepler-210b & WASP-19b & HAT-P-11b\\
\hline
\textbf{Mass ($M_{\mathrm{J}}$)} & $ < 0.377$ & $\sim$0.05 & $\sim$0.08 & 1.139$\pm$0.036 & 0.084$\pm$0.007\\
\textbf{Radius ($R_{\mathrm{J}}$)} & $ 0.540 \pm 0.020$  & $0.436 \pm 0.004$ & $0.342 \pm 0.003$ & 1.410$\pm$0.021 & 0.396$\pm$0.009\\  
\textbf{Orbital Period (days)} & 9.434$\pm$0.001 & 7.973$\pm$0.001& 2.453$\pm$0.001 & 0.7888 & 4.887\\
\textbf{Semi-major axis (AU)} & 0.080$\pm$0.002 & 0.0371$\pm$0.0010 & 0.0142$\pm$0.0002 &0.0165 & 0.0530\\
\textbf{Mean density ($\rho$ in g cm$^{-3}$)}&$<3.2$&$<$0.81&$<$2.67&0.542$\pm$0.030&1.806$\pm$0.194\\
\hline
Observation log&&&&\\
\hline
\textbf{Observation start} & 2014-09-28 11:35:03 & 2014-10-15 01:06:32 & & 2014-07-10 12:39:10 & 2015-05-19 12:48:40\\
\textbf{Observation end}  & 2014-09-28 19:42:01 & 2014-10-15 07:06:54 &  & 2014-07-10 18:27:45 & 2015-05-19 21:15:35\\
\textbf{Duration (ks)} & 28.67 & 19.73 & &20.45 & 27.00\\
\textbf{Filter} & Thin &Thin & &Thin &Thin \\
\textbf{ObsID} & 0743460301 & 0743460201 & &  0743460501 & 0764100701\\
\hline
\end{tabular}

\footnotesize{\textbf{References:} Stellar parameters for Kepler-63 (except spectral type) and its planetary companion parameters are taken from \cite{ojeda_2013}; the spectral type is obtained from exoplanetkyoto.org. Stellar parameters and planetary companion parameters for Kepler-210 are taken from \cite{ioannidis2014kepler}. The masses of Kepler-210 b and c are estimated using the mass-radius relationship given by \cite{lissauer2011architecture}. For WASP-19 and its planetary companion, all parameters are from \cite{hebb_2010}. For HAT-P-11 and its planetary companion, parameters are taken from \cite{bakos_2010}. The mean planetary densities are used to infer the composition and internal structure of orbiting exoplanets \citep{baraffe_2008,fortney_2010,swift_2012,spiegel_2014}. }
\end{table*}

\subsection{Kepler-63}

Kepler-63 is a young Sun-like star ($\sim$0.21$\pm$0.04 Gyr, \citealt{ojeda_2013, estrela_2016}) at a distance of 
$200 \pm 15$ parsec with a mass of 0.98M$_{\odot}$, effective temperature of 5576 K \citep{ojeda_2013} and a rotation period of 5.4 days \citep{estrela_2016}. 
The data presented by \cite{ojeda_2013} suggest 
that Kepler-63 has a rather high level of chromospheric activity (measured in Ca~II~H\&K)
with $\log$ R$\mathrm{ '_{HK}}=-4.39$;
one also observes quasi-periodic stellar flux variations on the
order of  $\sim 4\%$, interpreted as rotational modulation by starspots.  Furthermore, based on spot 
modeling from the {\it Kepler} data, \cite{estrela_2016} argue for the existence of
a magnetic activity cycle of 1.27$\pm$0.16 years in Kepler-63.

Kepler-63 hosts a hot Jupiter with a radius of $6.1\pm 0.2$ R$_{\oplus}$ in a polar orbit with 
a relatively short orbital period of 9.434 days and a semi-major axis of 0.08 AU \citep{ojeda_2013}. 
The mass of the planet is not known very precisely because its radial velocity data are 
contaminated by the 
high activity level of the host star. However, the RV data suggest a semi-amplitude value of 
approximately 
15-20~m/s, which leads to an upper limit of 120 M$_{\oplus}$  for the mass of Kepler-63b.
\citep{ojeda_2013}. 
Based on these values for mass and radius, the upper limit for the planet's mean
density is $\rho_p<3.0$ g/cm$^{3}$ \citep{ojeda_2013}.

\begin{figure*}
\begin{center}
\includegraphics[width=0.49\textwidth,clip]{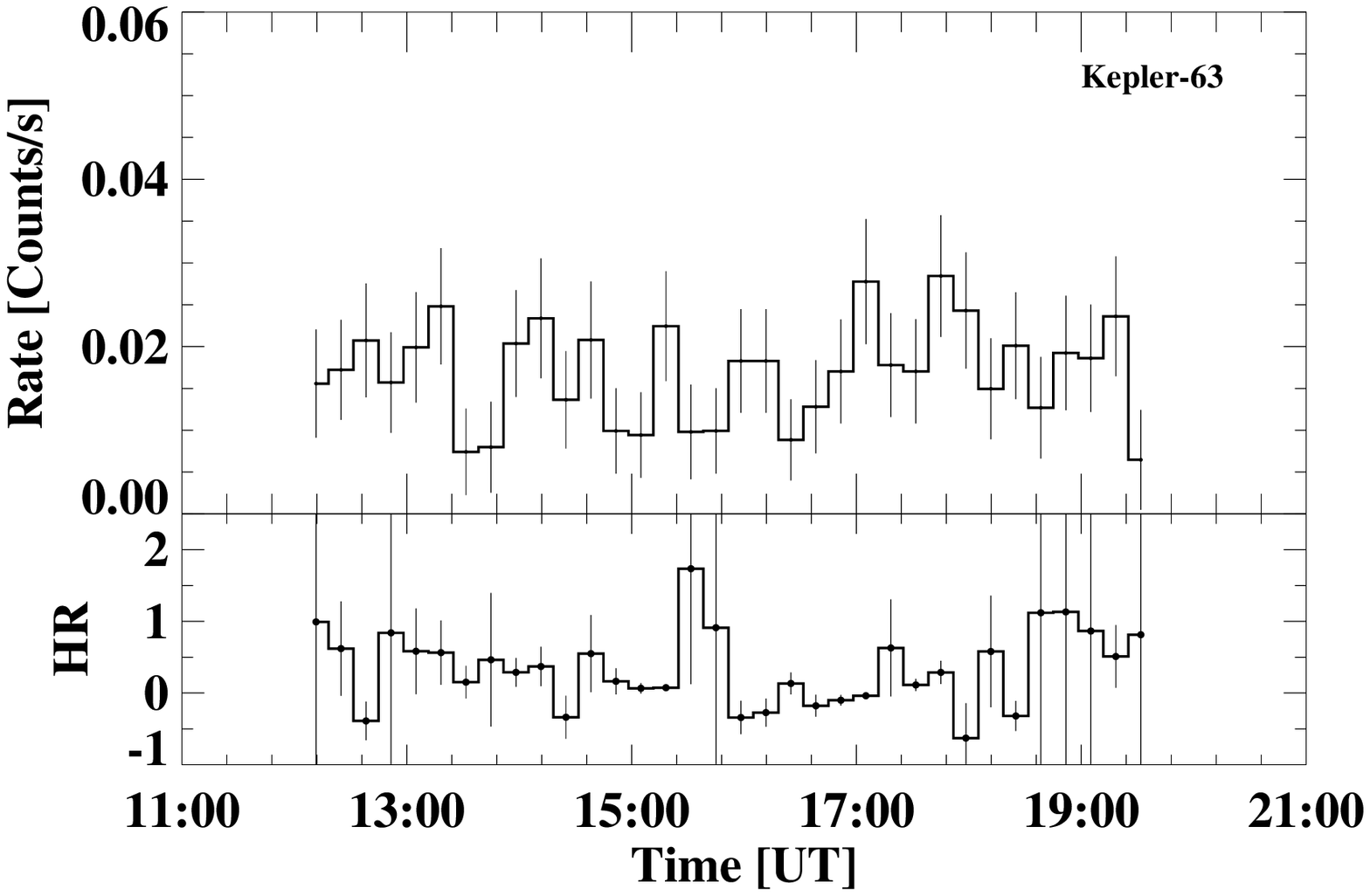}\vspace{-0.4mm}
\includegraphics[width=0.49\textwidth,clip]{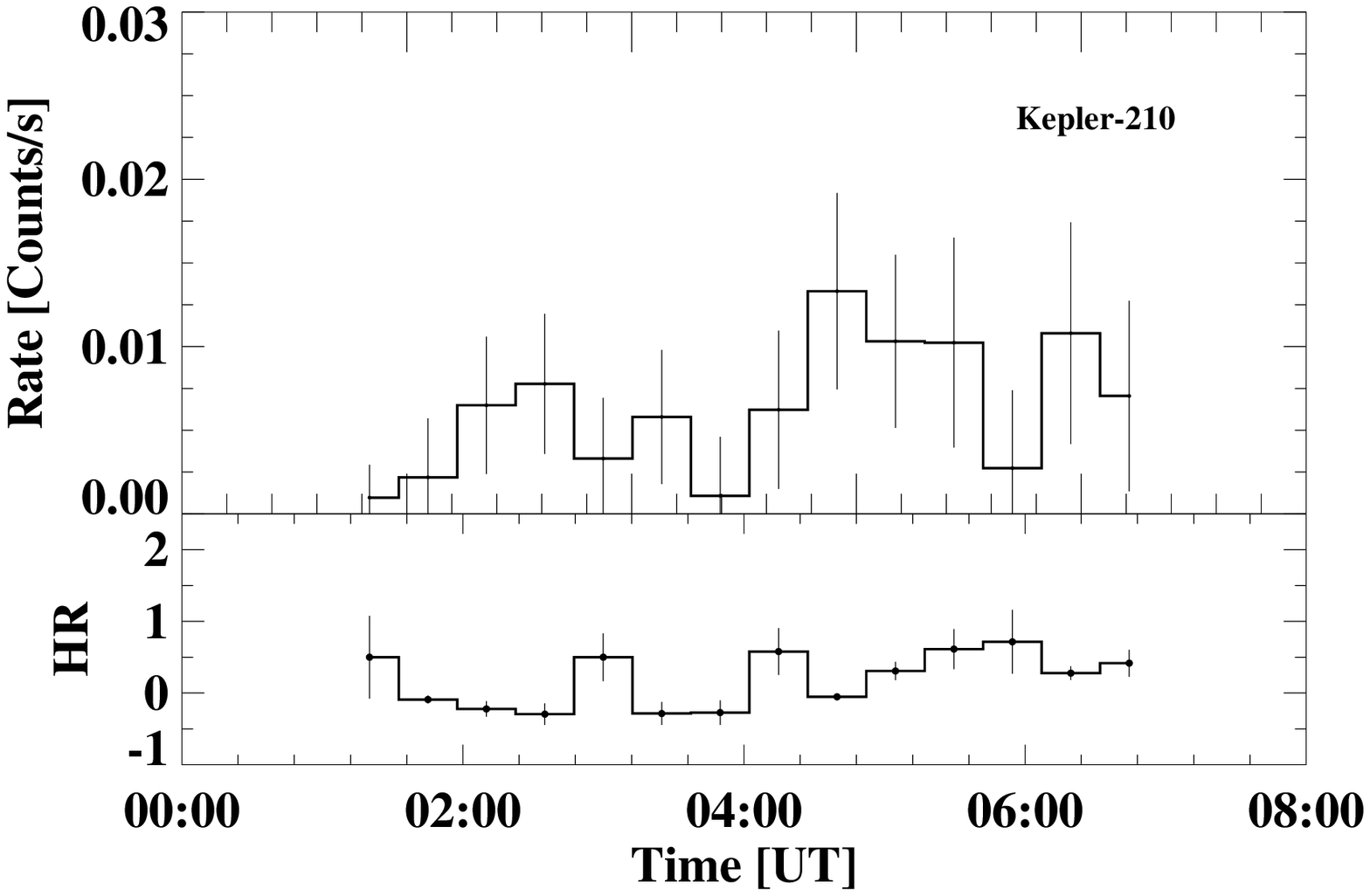}\vspace{-0.4mm}
\includegraphics[width=0.49\textwidth,clip]{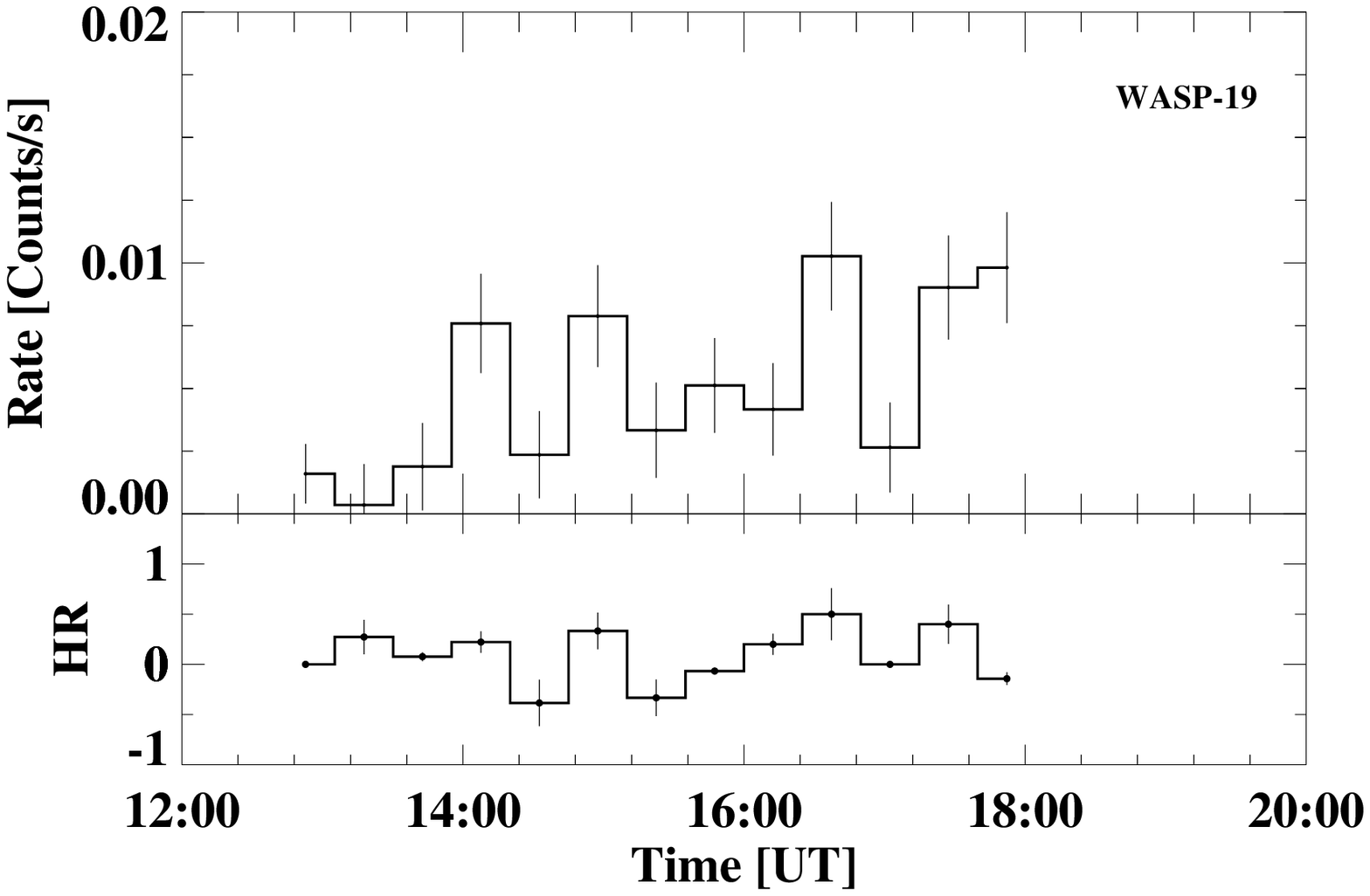}\vspace{-0.4mm}
\includegraphics[width=0.49\textwidth,clip]{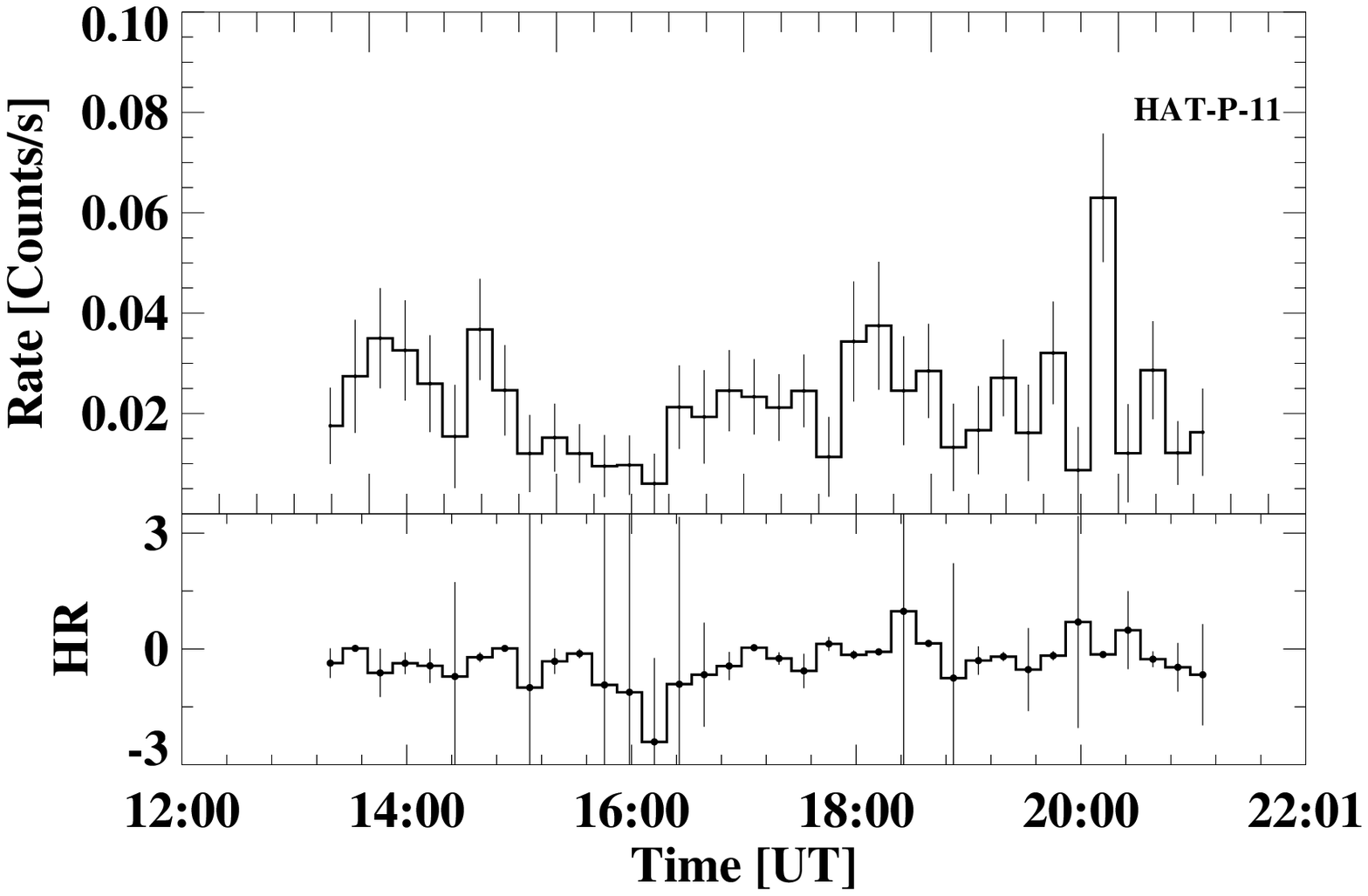}\vspace{-0.4mm}
\caption{\label{light_curve1} Light curves of Kepler-63, Kepler-210, WASP-19 and HAT-P-11 observed by EPIC detectors. The hardness ratio is plotted in lower panel.}
\end{center}
\end{figure*}

\subsection{Kepler-210}

Kepler-210 is a young active K-dwarf with a period of 12.33 days and an estimated
age of 350$\pm$50 Myr \citep{ioannidis2014kepler}; the activity  of Kepler-210 is shown by the large
modulations in the observed {\it Kepler} light curve.
Kepler-210 hosts at least two Neptune-sized planets, Kepler-210b with a 
radius of $3.75 \pm 0.03$ R$_{\oplus}$ and
a period of 2.453 days, and Kepler-210c with a radius of $4.78 \pm 0.04$ R$_{\oplus}$ and 
a period of 7.972 days \citep{ioannidis2014kepler}.  According to \cite{ioannidis2014kepler}, these 
planets orbit at distances of 0.014~AU and 0.037~AU, respectively.  No
radial velocity data are available for the system and hence the masses of the planets are not known. 
However, using the simple mass-radius relationship 
$\mathrm{M_{p}} = \mathrm{\bigg(\dfrac{R_{p}}{R_{\oplus}}\bigg)^{2.06}M_{\oplus}}$ given by \cite{lissauer2011architecture}, which
holds reasonably well for planets smaller than Saturn, we may estimate the masses at 
least to within some factors 
and hence can provide mass loss estimates correct to an order of magnitude. 
Using this relation we find the masses of Kepler-210b and Kepler-210c to 
be 15.92~$\mathrm{M_{\oplus}}$ and 26.27~$\mathrm{M_{\oplus}}$ respectively, 
leading to mean density values of 1.66~$g/cm^{3}$ and 1.32~$g/cm^{3}$ respectively. \cite{ioannidis2014kepler} also argue that the densities 
of these planets are below 5 $g/cm^{3}$, with both planets having a mass  upper limit 
of 0.5 $\mathrm{M_{J}}$.

\subsection{WASP-19}

WASP-19 is an active G8V type star ($\log$ R$\mathrm{ '_{HK}}=-4.66$, \citealt{knutson_2010}) with a mass of 0.96~M$_{\odot}$, a radius of 0.94~R$_{\odot}$ and a 
surface gravity $\log$~g of 4.47$\pm$0.03 \citep{hebb_2010}. WASP-19 shows strong rotational modulation 
with a period of 10.5~days and,  according to the gyrochronology relationship given by \cite{barnes_2007}, this 
short period indicates that the system is relatively young ($\sim$600 Myr old).  
On the other hand, the estimated system age of $5.5^{+9.0}_{-4.5}$~Gyr  
derived by \cite{hebb_2010} using isochrone fitting and adopting a zero eccentricity value, is 
at least formally larger, yet the errors of this estimate are huge.

The orbiting exoplanet around WASP-19 is the largest objects in our sample with a radius of 15.21~R$_{\oplus}$ and a
mass of 371~M$_{\oplus}$, making it even larger than Jupiter.  WASP-19b has a short orbital period of 
only 0.7888~days, leading to a very small distance between host star and planet of 0.0162~AU. 
The mean density calculated from these mass and radius values is 0.58 $g/cm^{3}$ and  the mean density estimated by \cite{hebb_2010} 
is about 0.51$\pm$0.06~$g/cm^{3}$. However, it is interesting to note that the mean density of 
WASP-19b is just half that of Jupiter, suggesting that the 
planet could be slightly bloated for its mass. The enhanced radius of WASP-19b could 
be interpreted as a result of the high-energy irradiation from the host star or of
tidal energy dissipation.

\subsection{HAT-P-11}

HAT-P-11 is an active metal-rich K4 dwarf star ([Fe/H]=+0.31$\pm$0.05) located at a distance of 
38~pc. Using spectroscopic and photometric data, \cite{bakos_2010} estimated the mass, radius, temperature and age of HAT-P-11 to be 0.81~M$_{\odot}$, 0.75~R$_{\odot}$, 
4780$\pm$50~K and 6.5$^{+5.9}_{-4.1}$~Gyr respectively. Based on the Kepler photometric data, the stellar activity for HAT-P-11 is indicated by spot induced modulations \citep{ojeda_2011, deming_2011}. The star also shows strong chromospheric emission (measured in Ca~II H\&K) with an  S-index of 0.61 and $\log$ R$\mathrm{'_{HK}}=-4.585$ \citep{bakos_2010}, 
indicating again that the star is active. 

HAT-P-11 hosts a super-Neptune with a radius of 4.96~R$_{\oplus}$, a 
mass of 25.74~M$_{\oplus}$ and a mean density of $\sim1.33$ ~$g/cm^{3}$, orbiting at a 
distance of 0.053~AU with a period of 4.887~days  \citep{bakos_2010}. 
When comparing to the models of \cite{fortney_2007}, \cite{bakos_2010} note
that, first,  the radius of HAT-P-11b appears to be much smaller than that of a
planet with similar mass and consisting of a 50\% ice/rock core and a 50\% H/He envelope;
and that, second, it is much larger than a planet with a pure 
ice/rock core without any H/He gaseous envelope.
Therefore \cite{bakos_2010} argue on the basis of
several parameters such as irradiation, distribution of heavy elements, age, etc., that
HAT-P-11b is more likely a super-Neptune 
planet with Z=0.9 and 10\% H/He envelope.

\section{Data analysis and results}

The data presented in this paper have been obtained with the \emph{XMM-Newton} 
satellite,  an observation log of our observation is contained in Tab.~\ref{tab:tab1}.
All X-ray data were reduced with the XMM-Newton Science Analysis System (SAS) software, version 15.0.0; EPIC light curves and spectra were obtained using standard filtering criteria and spectral analysis was carried out with XSPEC version 12.9.0  \citep{xspec}.

To provide an impression of the
data and their quality we show the soft X-ray images from the merged EPIC data and the extraction circles used for source and background counts (cf.,  Fig.~\ref{epicimg}).
Clearly, X-ray emission from all sample stars is detected, but with different strengths.
The source signals in case of Kepler-63 and HAT-P-11 were taken from a circular region with $12.5''$ radius with the co-ordinate position of the stars as the center. We measure the background signals from a circular region of $50''$ radius in a source-free region close to our targets. For Kepler-210, the revised coordinates are used as the center of a circular source extraction region with a radius of $12.5''$ and the background is taken from a circle with a radius of $40''$ far away from the source. 
For WASP-19, the source signal is from a circular region with a radius of $12.5''$ region 
with the coordinates of the candidate used as the center and the
background is estimated again from a circular region with a radius of $40''$.

\subsection{Temporal analysis}

As can be clearly seen in Fig.~\ref{epicimg}, the \emph{XMM-Newton} observations indicate obvious X-ray detections of all four host stars. 
Consequently, we carry out a more detailed temporal study of each of our targets and produce the X-ray light curves shown in Fig.~\ref{light_curve1}. 
The reduced PN data of Kepler-63 and HAT-P-11 in the 0.2-2.0 keV energy band are binned at 800~s. 
However, to enhance the signal we combine (in the cases of Kepler-210 and WASP-19) 
all the EPIC data and bin at 1500~s in the energy range of 0.2-2.0~keV. 
In the top panel of Fig.~\ref{light_curve1}, we show the source light curves, and in the bottom panel
the measured hardness ratio (HR) values vs. time; HR is defined as the fractional difference between hard energy band 
and soft energy band photon 
counts through the expression $ HR = \dfrac{H-S}{H+S}$, 
where $H$ is the number of counts between 0.7 and 2.0 keV (hard band) and S the number of counts between 0.2 and 0.7 keV (soft band). 
The mean $HR$-values determined from the PN data for  Kepler-63 and  HAT-P-11 are
$0.37\pm0.11$ and $0.34 \pm 0.30$ respectively, while those derived for Kepler-210 and WASP-19 
from the EPIC data are $0.22\pm0.27$ and  $0.12\pm0.16$, respectively.  The derived HR-values are consistent with each other to within the errors
and we observe no significant changes in count rate nor in HR, suggesting that the targets were observed  in a state of quiescence.

\subsection{X-Ray luminosities of exoplanet hosts}

\begin{figure}
\begin{center}
\includegraphics[width=0.49\textwidth,clip]{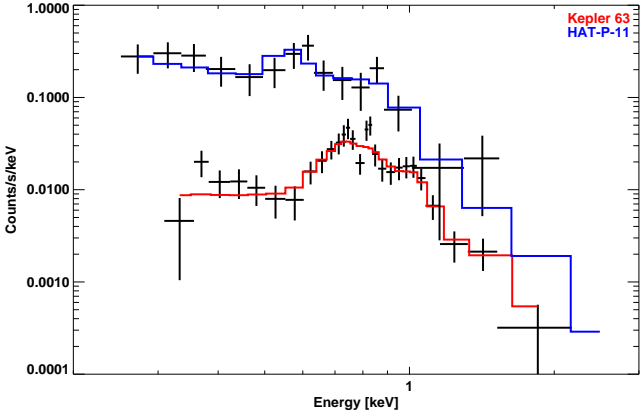}\vspace{-0.4mm}
\caption{\label{fig:spec} EPIC-PN spectra of Kepler-63 (red) and HAT-P-11 (blue) 
applied spectral models. HAT-P-11 spectra and the model are shifted by an order of 
magnitude for comparison.}
\end{center}
\end{figure}

For Kepler-63 and HAT-P-11, the number of counts measured in the MOS and pn cameras 
is sufficient for a more detailed spectral analysis.  In Fig.~\ref{fig:spec} we 
show the PN spectra for Kepler-63 and HAT-P-11 with their respective best-fit models;
the HAT-P-11 spectrum in Fig.~\ref{fig:spec} is shifted upwards by an order of magnitude
for better visibility.  A two-component APEC with solar abundance \citep{grevesse_1998}
provides a robust fit, the results of which are listed in Tab.~\ref{tab:specfit}, which also
gives the resulting X-ray fluxes and luminosities in the 0.2-2.0~keV band.  We also
experimented with treating the coronal abundances as a free parameter, but this did not 
significantly improve the fit and we conclude that the metallicity can be only poorly 
constrained with the available data.  Furthermore, since Kepler-63 is at a distance of 200 pc,
we also allow the absorption to vary freely and obtain an equivalent hydrogen column of 
$9.4^{+5.6}_{-6.2} \times10^{20}$ cm$^{-2}$; obviously the data quality is not sufficient to
constrain the amount of interstellar absorption along the line of sight towards Kepler-63 
very well. Inspection of Tab.~\ref{tab:specfit} shows 
that Kepler-63 is hotter and more X-ray luminous ($2\times10^{29}$~erg/s ) 
than  HAT-P-11 and in particular the Sun, indicating that Kepler-63 is indeed very active.

For Kepler-210 and WASP-19 we do not have sufficient counts for a 
meaningful spectral analysis, rather we use the HR-values to 
determine the coronal temperatures of the stars.
Using a two-component APEC (Astrophysical Plasma Emission Code) coronal plasma model for a collisionally 
ionised optically-thin thermal plasma \citep{smith_2001} with solar abundances, 
we calculate the theoretical HR values as a function of temperature. We performed the spectral analysis using XSPEC 
version 12.9.0 \citep{xspec}. We estimate the coronal temperature of Kepler-210 
to lie between 2.5 and 4.1 MK and that of WASP-19 between 3.3 and 5.3 MK. 
Using the estimated coronal temperatures in WebPIMMS, we convert the mean count rate measured with the 
XMM-Newton PN detector into an X-ray flux in the 0.2-2.0 keV energy band. 
Both Kepler-210 and WASP-19 are at  distances $>200$pc, where the interstellar hydrogen might produce a 
significant absorption of the X-ray flux.  
% TBD: The typical densities observed in the local cavity is $\sim0.1$cm$^{-3}$ \citep{cox_2005}. 
%The interstellar Na column density is a tracer of hydrogen \citep{ferlet_1985}, hence we used the catalog of 
%interstellar absorption provided by \cite{welsh_2010} to estimate the column density. 
We assume a hydrogen column depth of N(H$_1$ + H$_2$)$\sim$10$^{21}$cm$^{-2}$ with a canonical density of 1 particle per cm$^3$  \citep{ferlet_1985, cox_2005, welsh_2010}. We are aware that the hydrogen column density is not homogenous along all directions. 
%Hence, we also used the flux obtained Kepler-63 using spectral fitting as a conversion factor and estimate the fluxes for Kepler-210 and WASP-19. Both the  
%fluxes are comparable. 
Using a distance of 250 pc and 265 pc for WASP-19 and Kepler-210, respectively, 
the unabsorbed fluxes obtained using WebPimms are converted into the X-ray luminosities (see Tab.~\ref{tab:tab2}).

\begin{table}
\centering
\caption{Spectral modeling results for Kepler 63 and HAT-P-11 from the PN data.}
\label{tab:specfit}
\begin{tabular}{lcccc} 
\hline
 Parameters & Kepler-63 & HAT-P-11& Unit\\
 \hline
 T$_1$ &$0.43_{-0.08}^{+0.04}$&$0.14_{-0.10}^{+0.08}$&keV\\
 EM$_1$&$44.61_{-13.83}^{+22.28}$& $1.45_{-0.28}^{+0.21}$ &10$^{50}$cm$^{-3}$\\
 T$_2$ &$1.08^{+0.27}_{-0.22}$&$0.64_{-0.33}^{+0.23}$&keV\\
 EM$_2$&$14.09^{+4.80}_{-6.71}$& $0.35_{-0.09}^{+0.10}$ &10$^{50}$cm$^{-3}$\\
 N$_H$&$9.4^{+5.6}_{-6.2}$&--&10$^{20}$ cm$^{-2}$\\
\hline
$\chi^2_{\mathrm{red}}$(d.o.f)& 0.92 (25) & 0.87 (12)&\\
\hline
F$_{\mathrm{X}}$ (0.2-2.0 keV) & $2.35_{-1.19}^{+0.67}$&$2.77_{-0.12}^{+0.15}$&$10^{-14}$ erg/s/cm$^{-2}$\\
L$_{\mathrm{X}}$ (0.2-2.0 keV) & $10.06_{-2.67}^{+5.05}$ & $0.47_{-0.11}^{+0.20}$&$10^{28}$erg/s\\
\hline

\end{tabular}
\end{table}

\begin{table*}
\centering
\caption{Observed properties of the star candidates}
\label{tab:tab2}
\begin{tabular}{lccccc} 
\hline
 & \textbf{Kepler-63} & \textbf{Kepler-210} & \textbf{WASP-19} & \textbf{HAT-P-11}  \\ [0.5ex] 
\hline
Mean count rate [cts/ks]&14.47$\pm$0.98&3.14$\pm$0.76& 3.02$\pm$0.59&12.08$\pm$2.98\\
Average Coronal Temperature (MK) & 8.0 & 3.3 & 4.3 & 6.0\\ 
Flux [ $10^{-14}$ erg s$^{-1}$ cm$^{-2}$ in 0.2-2.0 keV] &2.35$_{-1.19}^{+0.67}$ &0.15$\pm$ 0.01 & 0.42$\pm$0.02  &2.77$_{-0.12}^{+0.15}$\\
$\log$ L$_{\mathrm{X}}$ in erg $s^{-1}$ & 29.00$_{-0.11}^{+0.21}$& 28.10$\pm$0.03 & 28.49$\pm$0.27 & 27.68$^{+0.03}_{-0.02}$ \\%&27\\
log L$_{\mathrm{bol}}$ in erg $s^{-1}$ & 33.41 & 33.14& 33.42& 33.25 \\%&\\
$\log$ ($\frac{\mathrm{L}_{\mathrm{X}}}{\mathrm{L}_{\mathrm{bol}}}$) & -4.36  & -5.03  & -4.93 & -5.57 \\%& -6.60 \\
\hline
$\log$ L$_{\mathrm{EUV}}$ in erg $s^{-1}$ &29.74 & 28.98& 29.31&28.60\\% & 28.02 \\
$\log$ L$_{\mathrm{XUV}}$ in erg $s^{-1}$ & 30.19& 29.03& 29.37&28.65 \\%& 28.06\\
$\log$ F$_{\mathrm{X}}$ at planet in erg $s^{-1}$$cm^{-2}$ & 3.74 & 4.36 (for b) and 3.53 (for c) &4.61 & 2.78 \\%& -0.45 (for Earth) and -1.88 (for Jupiter)\\
$\log$ F$_{\mathrm{XUV}}$ at planet in erg $s^{-1}$$cm^{-2}$ & 4.56 & 5.28 (for b) and 4.44 (for c) & 5.48& 3.75\\%& 0.61 (for Earth) and -0.82 (for Jupiter)\\
Upper-limit mass loss rates $\dot{\mathrm{M}}$ in $10^{11}$ g~s$^{-1}$ & 7.41 & 23.18 (for b) and 4.27 (for c)  & 63.17 &0.65\\
\hline
\end{tabular}
\end{table*}

The bolometric luminosity of our candidates can be calculated through the relation
$\mathrm{L_{\mathrm{bol}} = 10^{0.4(4.74-m_{v}-BC+5\log(d)-5)}L_{\odot}}$, where
$m_{\text v}$ denotes the apparent visual magnitude of the star, $BC$ the bolometric correction, $d$ the distance (in pc) and $L_{\odot}$ the solar bolometric luminosity. 
For Kepler-63, WASP-19 and HAT-P-11 the values for the V and K magnitudes are obtained from the SIMBAD database. 
For Kepler-210, we use the B-V color from \cite{ioannidis2014kepler} and convert it into a V-K color using 
the data provided in \cite{worthey2011empirical}; the bolometric corrections are determined using the relationship given by  \cite{worthey2011empirical}. 
The results of our analysis for all the targets are listed in Tab.~\ref{tab:tab2}.  
The computed $\log \mathrm{ \frac{L_X}{L_{bol}}}$-ratios indicate again stars of moderate activity with the
exception of Kepler-63. 
%The results of our analysis for all the targets are listed in  Tab.~\ref{tab:tab2}.

\subsection{Stellar radiation at the planetary positions}

As indicated at the end of the previous subsection, the $\log \mathrm{ \frac{L_X}{L_{bol}}}$ values for 
our sample stars lie between -4.36 and -5.57. 
Given these activity levels and the fact that the planets are close to their host stars, makes 
all of them likely to undergo atmospheric loss through hydrodynamic escape. 
As shown by \cite{lammer2003atmospheric}, such mass loss occurs predominantly 
due to radiation in the EUV and X-ray range.  From our XMM-Newton observations 
only X-ray wavelengths are accessible, hence we need to
apply  an indirect method to obtain the total high-energy radiation in the
X-ray and UV-bands. \cite{sanz2011estimation} compute the EUV spectra for a number 
of stars using the emission measure distribution derived from X-ray spectra. 
This approach has been shown to be quite accurate by further studies by \cite{claire2012evolution} and  
\cite{linsky2013intrinsic}. However, we do point out in this context that \cite{france2016} argue that FUV 
flux estimates based on soft X-ray emission alone can substantially underestimate 
the irradiation received by extrasolar planets in individual cases; since no detailed
FUX/XUV studies are available for our target stars, there is nothing we can do about this at the moment.
Using the scaling relations given by \cite{sanz2011estimation} we extrapolate the measured X-ray luminosity to the total high-energy radiation viz.
\begin{equation}\label{eqn:leuv}
	\log L_{\mathrm{EUV}}=(0.860\pm0.073)\log L_{X}+(4.80\pm1.99)
\end{equation}
and
\begin{equation} \label{eqn:lxuv}
	\log~L_{\mathrm{XUV}}=\log (L_{\mathrm{X}}+L_{\mathrm{EUV}}).
\end{equation}

\subsection{Estimated atmospheric mass loss of planetary candidates}

According to the hydrodynamic mass loss model as developed by \cite{watson1981dynamics}, \cite{lammer2003atmospheric}, 
\cite{sanz2010scenario} and \cite{erkaev2007roche}, the planetary mass loss 
rate $\dot{M}$ is given by
\begin{equation}\label{eqn:m}
	\dot{\mathrm{M}}=\dfrac{3 \pi \beta^{2} \epsilon F_{\mathrm{XUV}}}{4GK\rho_{\mathrm{p}}},
\end{equation}
where $\rho_{p}$ is the mean planetary density, $F_{XUV}$ the incident X-ray+EUV flux, and
G the gravitational constant. 
The term $\beta = R_{XUV}/R_{p}$ is a correction factor for the size of the planetary disk absorbing the XUV radiation. \cite{salz2016energy} provide an estimate for $\beta$ through
\begin{equation}\label{eqn:beta}
	\log\beta = max(0.0,-0.185 \ \log(-\phi_{G})+0.021 \ \log(F_{XUV})+2.42),
\end{equation}
where $\phi_{G}=-GM_{p}/R_{p}$ is the gravitational potential at the surface of the planet.
We assume an heating efficiency $\epsilon = 0.4$ (in Eqn.~\ref{eqn:m}) as suggested by \cite{valencia2010composition} for hot Jupiters 
and strongly irradiated rocky planets. 
Several authors choose values of  $\epsilon<0.4$ \citep{owen_2012, shematovich_2014},
based on observations of evaporating hot Jupiter HD~209458b. 
Furthermore, \cite{salz2016energy} show that small planets like HAT-P-11 do not show trend in the 
heating/evaporating efficiency and hence using 
a constant value of $\epsilon$ is a reasonable approximation. 
Note that the value of $\epsilon$ is only a rough estimate, however, in reality, the 
heating efficiency can be determined more precisely only after modeling the atmosphere 
of exoplanets in detail. 

The parameter K in Eqn. \ref{eqn:m} is a factor which takes 
into account the effects of mass loss through Roche lobe overflow.  
According to  \cite{erkaev2007roche},  K is given by 
\begin{equation}\label{eqn:keta}
	K(\eta)=1-\dfrac{3}{2\eta}+\dfrac{1}{2\eta^{3}},
\end{equation}
where the parameter $\eta$ is given by
\begin{equation}\label{eqn:eta}
	\eta = \bigg(\dfrac{M_{p}}{3M_{s}}\bigg)^{1/3}\dfrac{a}{R_{p}}.
\end{equation}
Here $M_{p}$ and $M_{s}$ denote planetary and stellar masses respectively, $a$ is the semi-major axis and 
$R_{p}$ the planetary radius. 
This expression is approximate and valid only if $a>>R_{Rl}>R_{p}$ (where $R_{Rl}$ is the Roche lobe radius of the planet) and $M_{s}>M_{p}$ hold. 
Following \cite{erkaev2007roche} and taking $R_{Rl}$ as 
\begin{equation}\label{eqn:rrl}
	R_{Rl} = a\bigg(\dfrac{M_{p}}{3M_{s}}\bigg)^{1/3},
\end{equation}
we find $\eta = R_{Rl}/R_P $.  Since $R_{Rl}$  is usually much larger than the radius Rp, $K(\eta)$ approaches unity.

%\cite{seager_2007} show that the Roche lobe radius exceeds 
%the initial exoplanetary radius, if the initial mass of the exoplanet is three times its . 
%Hence the Roche lobe influenced mass loss of our exoplanets is likely insignificant. 

\begin{figure}
		\begin{center}
		\includegraphics[width=0.47\textwidth]{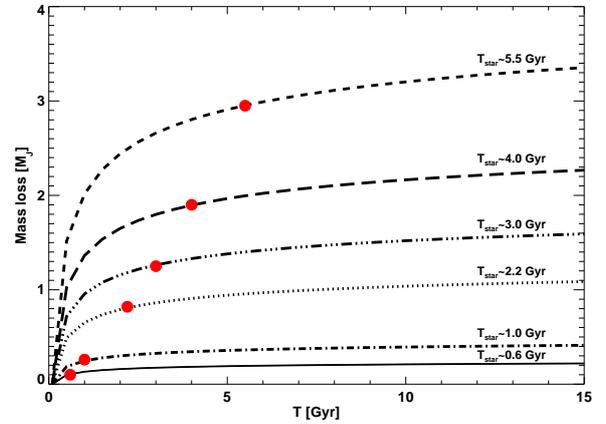}
		\caption{\label{ageevol} Evolution of the total planetary mass loss of WASP-19 for the current X-ray luminosity. The current ages are indicated as filled circles.}
		\end{center}
\end{figure}

With the above equations, the planetary parameters listed in Tab.~\ref{tab:tab1} and the observed
X-ray luminosity of the planet host stars, the upper-limit mass loss rates can be estimated. We note that the mass-loss rates are derived 
assuming a constant density. The mass loss induced by the incident high-energy radiation 
will in general also change the radius of the planet, since mass and radius are coupled,
once the composition and equation of state for the interior are known.
\cite{lopez_2014} studied in detail the change in the radius of the planet as a function of its 
mass for different planetary compositions and in particular for different contributions 
of a H/He envelope.  Since the compositions of the planets studied in this paper are not known, 
we assume the simplest model of a constant mean density of the planet to estimate the 
mass-loss history. We also note that the changing incident XUV flux provides the by far dominant
component for driving the mass loss.

Stellar activity, especially for young stars is not constant over the lifetimes of the stars. In fact, according to \cite{ribas2005evolution} the X-ray luminosity varies approximately 
by a factor of $\bigg(\dfrac{\tau}{\tau_{\star}}\bigg)^{-1.23}$, where  $\tau$ is the current stellar 
age and $\tau_{\star}$ is the stellar age, at which stellar activity remains at a constant level (at about 0.1~Gyr),
which is also taken as the start time for the integration of the mass loss. 
For single stars, the coronal activity is highest when the star 
has --~more or less~-- finished
its contraction, but has not yet spun down, so that the large stellar 
rotation yields very high levels of activity \citep{sanz_2014}; hence we choose as starting time
an age of $0.1$ Gyr. 
Inserting this time variable 
flux into the hydrodynamic mass loss equation and integrating it over time, provides a coarse estimate
of the mass loss history of these planets. 
The results 
are listed in Table~\ref{tab:tab2}.

According to our estimates, both Kepler-210 b and c 
have suffered very little mass loss below $0.1$M$_J$. However, 
Kepler-63 b has suffered an mass loss of possibly up to 0.5 M$_J$.  
The age-activity relationship for stars older than 1 Gyr are steeper and are given by 
the relation  
$\log \mathrm{{ \frac{L_X}{(R_{\star}/R_{\odot})^2}} = 54.65\pm6.98 - (2.80\pm0.72) \log t}$ \citep{booth_2017}.
We calculate the maximal mass loss for HAT-P-11 b to lie in the range between 0.15-0.25M$_J$, 
using both the age-activity 
relations given by \cite{ribas2005evolution} and \cite{booth_2017}. 
Furthermore, the large uncertainity on the age of HAT-P-11 which 
can contribute significantly to estimated mass-loss of the orbiting planet. 

Similar to HAT-P-11, there are huge uncertainties 
in the proposed ages of WASP-19 which ranges from under 1~Gyr to more than 5~Gyr. In Figure~\ref{ageevol}, we 
plot the total mass lost integrating over stellar ages between 0.6-5.5 Gyr,  where
each curve represents a different age of the WASP-19 systems. 
Considering the large uncertainty in the age of WASP-19, we estimate a total mass loss of 0.1-3 M$_J$. However, assuming an age  of WASP-19 of $\sim$2.2~Gyr \citep{brown_2011} by comparing the X-ray luminosity of WASP-19 with the age-X-ray luminosity relation, we estimate the maximal mass lost by 
the planet to be between $\sim$0.9-1.0 M$_J$.

\begin{figure}
		\begin{center}
		\includegraphics[width=0.47\textwidth]{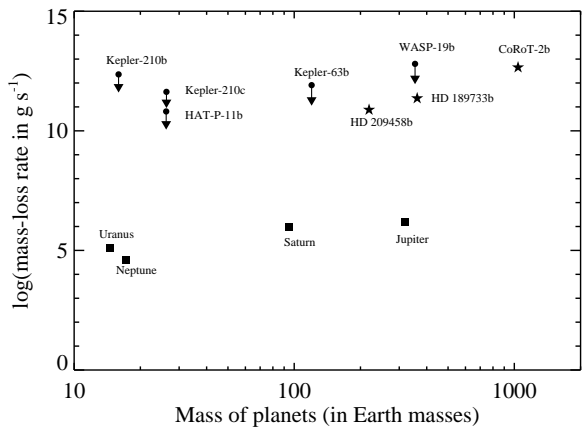}
		\caption{\label{compare_ss} $\log\dot{M}$ vs. Mass for solar system gaseous giant 
planets  (filled squares) and our sample exoplanets (filled circle with downward arrow).  The
three well-studied hot-Jupiter systems HD 189733, HD 209458 and CoRoT-2 are additionally shown as filled stars. 
All values of $\log\dot{M}$ are upper limit estimates using $\epsilon = 0.4$ and assuming $K=1$ without 
any magnetosphere. Using the mass-radius relationship given by  \citealt{lissauer2011architecture}, we estimate the masses of Kepler-210 b and c.}
		\end{center}
\end{figure}

\section{Discussion and Conclusions}

Here we report the results of exploratory X-ray studies of the four transiting active exoplanet
hosts Kepler-63, Kepler-210, WASP-19 and HAT-P-11 in the soft X-ray energy band 0.2~-~2.0~keV, which yield detection with logarithmic X-ray luminosities  in the range from 27.68-29.05 erg s$^{-1}$.
The \emph{XMM-Newton} pointing were rather short and allow only a rather coarse characterization of the
X-ray properties of the host stars.  All the target stars show an X-ray activity indicator 
$\log \mathrm{ \frac{L_X}{L_{bol}}}$ in the range between -4.36 to -5.57, which combined with coronal temperatures in the range 3.0 and 8.0 MK, 
points to moderately active to active stars. Our analysis suggests that the X-ray properties of Kepler-63 
are compatible with the presence of strong Ca~II H\&K emission cores and spot modulation. 
Using the measured X-ray luminosities, we infer the irradiation fluxes at far UV wavelengths 
allowing us to estimate the XUV flux at the planetary position and the corresponding mass-loss rate. 

To provide a comparison of how strongly the hot-Jupiter planets lose mass by stellar irradiation 
we also calculate mass loss estimates for solar system planets making the same assumptions as for
the extrasolar planets; this implies in particular that 
they have a hypothetical atmosphere of gases with mainly H/He (which is clearly not the case for the 
inner solar system planets). 
In Fig.~\ref{compare_ss} we plot the thus derived maximal mass loss rates for our
sample stars and those derived for the solar system gas giants.  Fig.~\ref{compare_ss} indicates all our 
exoplanetary candidates have maximal mass-loss rates some orders of magnitude larger than the
mass-loss rates of solar system gas giants; this is due to the fact that these planets are, first, 
orbiting much closer to their host stars, and second, that the intrinsic
X-ray luminosity of the hosts is much larger than that of the present-day Sun.

Furthermore,  we compare our four targets with two other well studied hot 
Jupiters HD~209458~b \citep{charbonneau1999detection} and HD~189733 b \citep{bouchy2005elodie}. 
HD~209458~b is an exoplanet with a mass of 0.69 $M_{J}$ and a radius of 1.38 $R_{J}$ orbiting a 
G0V star (46~pc away from the Sun) at a distance of 0.047 AU with a period of 3.52 days.  
The X-ray luminosity of HD~209458 has recently been measured by \emph{Chandra} HRC as
$\log$ L$_{\mathrm{X}}$ $\sim 27.20$ erg/s \citep{czesla_2017},
yielding an XUV luminosity of $\log$ L$_{\mathrm{XUV}} \sim$28.23 erg/s; it also shows ongoing mass loss through 
hydrodynamic escape \citep{sanz2010scenario}.  
HD~189733 b has a mass of 1.142~$M_{J}$ and radius 1.138 $R_{J}$ orbiting around a K1/K2 star HD~189733 (19~pc away from the Sun)
orbiting at a distance of 0.03 AU with a period of about 2.22 days. 
The X-ray and XUV of luminosities of HD~189733 are $\log$ L$_{\mathrm{X}} \sim 28.18$ and $\log$ L$_{\mathrm{X}} \sim 28.85$ erg/s, respectively \citep{poppenhaeger2013transit}.

All our targets are as bright as HD~209458 and HD~189733 in the X-ray and XUV ranges, 
making the planets extremely susceptible to atmospheric mass loss. 
Using these XUV luminosities, the mass loss rate of HD~209458 b is calculated to be  $9.3\times10^{10}$ gs$^{-1}$, 
consistent with the lower limit mass loss rate of  $7.6\times10^{10}$ gs$^{-1}$ derived by \cite{linsky2010observations}. 
For HD~189733~b, the mass loss rate was estimated  to be about $2.3\times10^{11}$ gs$^{-1}$ by 
\cite{poppenhaeger2013transit}, it is thus 
comparable to within an order of magnitude of the value obtained for Kepler-210c and Kepler-63b. 
In addition, we note that the mass loss rate of HD209458 b is comparable to the upper-limit mass loss rate of HAT-P-11b, 
although the latter orbits further away from its host star.  Also,
the upper-limit mass loss rates of  WASP-19 b and Kepler-210 b are at least an order of magnitude higher than those in the 
prototypical targets HD~209458 and HD~189733.  Rather, 
the mass loss rate of these planets are comparable to CoRoT-2b, which orbits around the extremely
active planet host star with an X-ray luminosity of $2\times10^{29}$ erg/s \citep{schroeter_2011}. 
The X-ray flux received by  CoRoT-2 b is $\sim9 \times10^4$erg cm$^{-2}$ s$^{-1}$, which corresponds 
to a formal mass loss rate of $4.5\times10^{12}$ g s$^{-1}$. However,
\cite{salz_2016a} show that CoRoT-2 is one of the most compact planets with very efficient radiative
cooling, thus preventing the development of a strong wind.  With the observations present here
we have laid the groundwork for further detailed studies of these exoplanet hosts.

\bibliographystyle{mnras}
\bibliography{paper} 
\begin{thebibliography}{}
\makeatletter
\relax
\def\mn@urlcharsother{\let\do\@makeother \do\$\do\&\do\#\do\^\do\_\do\%\do\~}
\def\mn@doi{\begingroup\mn@urlcharsother \@ifnextchar [ {\mn@doi@}
  {\mn@doi@[]}}
\def\mn@doi@[#1]#2{\def\@tempa{#1}\ifx\@tempa\@empty \href
  {http://dx.doi.org/#2} {doi:#2}\else \href {http://dx.doi.org/#2} {#1}\fi
  \endgroup}
\def\mn@eprint#1#2{\mn@eprint@#1:#2::\@nil}
\def\mn@eprint@arXiv#1{\href {http://arxiv.org/abs/#1} {{\tt arXiv:#1}}}
\def\mn@eprint@dblp#1{\href {http://dblp.uni-trier.de/rec/bibtex/#1.xml}
  {dblp:#1}}
\def\mn@eprint@#1:#2:#3:#4\@nil{\def\@tempa {#1}\def\@tempb {#2}\def\@tempc
  {#3}\ifx \@tempc \@empty \let \@tempc \@tempb \let \@tempb \@tempa \fi \ifx
  \@tempb \@empty \def\@tempb {arXiv}\fi \@ifundefined
  {mn@eprint@\@tempb}{\@tempb:\@tempc}{\expandafter \expandafter \csname
  mn@eprint@\@tempb\endcsname \expandafter{\@tempc}}}

\bibitem[\protect\citeauthoryear{{Arnaud}}{{Arnaud}}{1996}]{xspec}
{Arnaud} K.~A.,  1996, in {Jacoby} G.~H.,  {Barnes} J.,  eds, ASP Conf. Ser.
  101: Astronomical Data Analysis Software and Systems V. pp 17--+

\bibitem[\protect\citeauthoryear{{Bakos} et~al.,}{{Bakos}
  et~al.}{2010}]{bakos_2010}
{Bakos} G.~{\'A}.,  et~al., 2010, \mn@doi [\apj]
  {10.1088/0004-637X/710/2/1724}, \href
  {http://adsabs.harvard.edu/abs/2010ApJ...710.1724B} {710, 1724}

\bibitem[\protect\citeauthoryear{{Baraffe}, {Chabrier}  \& {Barman}}{{Baraffe}
  et~al.}{2008}]{baraffe_2008}
{Baraffe} I.,  {Chabrier} G.,   {Barman} T.,  2008, \mn@doi [\aap]
  {10.1051/0004-6361:20079321}, \href
  {http://adsabs.harvard.edu/abs/2008A%26A...482..315B} {482, 315}

\bibitem[\protect\citeauthoryear{{Barnes}}{{Barnes}}{2007}]{barnes_2007}
{Barnes} S.~A.,  2007, \mn@doi [\apj] {10.1086/519295}, \href
  {http://adsabs.harvard.edu/abs/2007ApJ...669.1167B} {669, 1167}

\bibitem[\protect\citeauthoryear{{Booth}, {Poppenhaeger}, {Watson}, {Silva
  Aguirre}  \& {Wolk}}{{Booth} et~al.}{2017}]{booth_2017}
{Booth} R.~S.,  {Poppenhaeger} K.,  {Watson} C.~A.,  {Silva Aguirre} V.,
  {Wolk} S.~J.,  2017, preprint, \href
  {http://adsabs.harvard.edu/abs/2017arXiv170608979B} {} (\mn@eprint {arXiv}
  {1706.08979})

\bibitem[\protect\citeauthoryear{Bouchy et~al.,}{Bouchy
  et~al.}{2005}]{bouchy2005elodie}
Bouchy F.,  et~al., 2005, Astronomy \& Astrophysics, 444, L15

\bibitem[\protect\citeauthoryear{{Brown}, {Collier Cameron}, {Hall}, {Hebb}  \&
  {Smalley}}{{Brown} et~al.}{2011}]{brown_2011}
{Brown} D.~J.~A.,  {Collier Cameron} A.,  {Hall} C.,  {Hebb} L.,   {Smalley}
  B.,  2011, \mn@doi [\mnras] {10.1111/j.1365-2966.2011.18729.x}, \href
  {http://adsabs.harvard.edu/abs/2011MNRAS.415..605B} {415, 605}

\bibitem[\protect\citeauthoryear{Charbonneau, Brown, Latham  \&
  Mayor}{Charbonneau et~al.}{1999}]{charbonneau1999detection}
Charbonneau D.,  Brown T.~M.,  Latham D.~W.,   Mayor M.,  1999, The
  Astrophysical Journal Letters, 529, L45

\bibitem[\protect\citeauthoryear{Charbonneau, Brown, Noyes  \&
  Gilliland}{Charbonneau et~al.}{2002}]{charbonneau2002detection}
Charbonneau D.,  Brown T.~M.,  Noyes R.~W.,   Gilliland R.~L.,  2002, The
  Astrophysical Journal, 568, 377

\bibitem[\protect\citeauthoryear{Claire, Sheets, Cohen, Ribas, Meadows  \&
  Catling}{Claire et~al.}{2012}]{claire2012evolution}
Claire M.~W.,  Sheets J.,  Cohen M.,  Ribas I.,  Meadows V.~S.,   Catling
  D.~C.,  2012, The Astrophysical Journal, 757, 95

\bibitem[\protect\citeauthoryear{{Cox}}{{Cox}}{2005}]{cox_2005}
{Cox} D.~P.,  2005, \mn@doi [\araa] {10.1146/annurev.astro.43.072103.150615},
  \href {http://adsabs.harvard.edu/abs/2005ARA%26A..43..337C} {43, 337}

\bibitem[\protect\citeauthoryear{{Czesla}, {Salz}, {Schneider}, {Mittag}  \&
  {Schmitt}}{{Czesla} et~al.}{2017}]{czesla_2017}
{Czesla} S.,  {Salz} M.,  {Schneider} P.~C.,  {Mittag} M.,   {Schmitt}
  J.~H.~M.~M.,  2017, preprint, \href
  {http://adsabs.harvard.edu/abs/2017arXiv170804537C} {} (\mn@eprint {arXiv}
  {1708.04537})

\bibitem[\protect\citeauthoryear{{Deming} et~al.,}{{Deming}
  et~al.}{2011}]{deming_2011}
{Deming} D.,  et~al., 2011, \mn@doi [\apj] {10.1088/0004-637X/740/1/33}, \href
  {http://adsabs.harvard.edu/abs/2011ApJ...740...33D} {740, 33}

\bibitem[\protect\citeauthoryear{Erkaev, Kulikov, Lammer, Selsis, Langmayr,
  Jaritz  \& Biernat}{Erkaev et~al.}{2007}]{erkaev2007roche}
Erkaev N.,  Kulikov Y.~N.,  Lammer H.,  Selsis F.,  Langmayr D.,  Jaritz G.,
  Biernat H.,  2007, Astronomy \& Astrophysics, 472, 329

\bibitem[\protect\citeauthoryear{{Estrela} \& {Valio}}{{Estrela} \&
  {Valio}}{2016}]{estrela_2016}
{Estrela} R.,  {Valio} A.,  2016, \mn@doi [\apj] {10.3847/0004-637X/831/1/57},
  \href {http://adsabs.harvard.edu/abs/2016ApJ...831...57E} {831, 57}

\bibitem[\protect\citeauthoryear{{Ferlet}, {Vidal-Madjar}  \& {Gry}}{{Ferlet}
  et~al.}{1985}]{ferlet_1985}
{Ferlet} R.,  {Vidal-Madjar} A.,   {Gry} C.,  1985, \mn@doi [\apj]
  {10.1086/163666}, \href {http://adsabs.harvard.edu/abs/1985ApJ...298..838F}
  {298, 838}

\bibitem[\protect\citeauthoryear{{Fortney} \& {Nettelmann}}{{Fortney} \&
  {Nettelmann}}{2010}]{fortney_2010}
{Fortney} J.~J.,  {Nettelmann} N.,  2010, \mn@doi [\ssr]
  {10.1007/s11214-009-9582-x}, \href
  {http://adsabs.harvard.edu/abs/2010SSRv..152..423F} {152, 423}

\bibitem[\protect\citeauthoryear{{Fortney}, {Marley}  \& {Barnes}}{{Fortney}
  et~al.}{2007}]{fortney_2007}
{Fortney} J.~J.,  {Marley} M.~S.,   {Barnes} J.~W.,  2007, \mn@doi [\apj]
  {10.1086/512120}, \href {http://adsabs.harvard.edu/abs/2007ApJ...659.1661F}
  {659, 1661}

\bibitem[\protect\citeauthoryear{{France} et~al.,}{{France}
  et~al.}{2016}]{france2016}
{France} K.,  et~al., 2016, \mn@doi [\apj] {10.3847/0004-637X/820/2/89}, \href
  {http://adsabs.harvard.edu/abs/2016ApJ...820...89F} {820, 89}

\bibitem[\protect\citeauthoryear{Gillon et~al.,}{Gillon
  et~al.}{2017}]{gillon2017seven}
Gillon M.,  et~al., 2017, Nature, 542, 456

\bibitem[\protect\citeauthoryear{{Grevesse} \& {Sauval}}{{Grevesse} \&
  {Sauval}}{1998}]{grevesse_1998}
{Grevesse} N.,  {Sauval} A.~J.,  1998, \ssr, 85, 161

\bibitem[\protect\citeauthoryear{{Hebb} et~al.,}{{Hebb}
  et~al.}{2010}]{hebb_2010}
{Hebb} L.,  et~al., 2010, \mn@doi [\apj] {10.1088/0004-637X/708/1/224}, \href
  {http://adsabs.harvard.edu/abs/2010ApJ...708..224H} {708, 224}

\bibitem[\protect\citeauthoryear{Ioannidis, Schmitt, Avdellidou, von Essen  \&
  Agol}{Ioannidis et~al.}{2014}]{ioannidis2014kepler}
Ioannidis P.,  Schmitt J.,  Avdellidou C.,  von Essen C.,   Agol E.,  2014,
  Astronomy \& Astrophysics, 564, A33

\bibitem[\protect\citeauthoryear{{Knutson}, {Howard}  \& {Isaacson}}{{Knutson}
  et~al.}{2010}]{knutson_2010}
{Knutson} H.~A.,  {Howard} A.~W.,   {Isaacson} H.,  2010, \mn@doi [\apj]
  {10.1088/0004-637X/720/2/1569}, \href
  {http://adsabs.harvard.edu/abs/2010ApJ...720.1569K} {720, 1569}

\bibitem[\protect\citeauthoryear{Lammer, Selsis, Ribas, Guinan, Bauer  \&
  Weiss}{Lammer et~al.}{2003}]{lammer2003atmospheric}
Lammer H.,  Selsis F.,  Ribas I.,  Guinan E.,  Bauer S.,   Weiss W.,  2003, The
  Astrophysical Journal Letters, 598, L121

\bibitem[\protect\citeauthoryear{Linsky, Yang, France, Froning, Green, Stocke
  \& Osterman}{Linsky et~al.}{2010}]{linsky2010observations}
Linsky J.~L.,  Yang H.,  France K.,  Froning C.~S.,  Green J.~C.,  Stocke
  J.~T.,   Osterman S.~N.,  2010, The Astrophysical Journal, 717, 1291

\bibitem[\protect\citeauthoryear{Linsky, Fontenla  \& France}{Linsky
  et~al.}{2013}]{linsky2013intrinsic}
Linsky J.~L.,  Fontenla J.,   France K.,  2013, The Astrophysical Journal, 780,
  61

\bibitem[\protect\citeauthoryear{Lissauer et~al.,}{Lissauer
  et~al.}{2011}]{lissauer2011architecture}
Lissauer J.~J.,  et~al., 2011, The Astrophysical Journal Supplement Series,
  197, 8

\bibitem[\protect\citeauthoryear{{Lopez} \& {Fortney}}{{Lopez} \&
  {Fortney}}{2014}]{lopez_2014}
{Lopez} E.~D.,  {Fortney} J.~J.,  2014, \mn@doi [\apj]
  {10.1088/0004-637X/792/1/1}, \href
  {http://adsabs.harvard.edu/abs/2014ApJ...792....1L} {792, 1}

\bibitem[\protect\citeauthoryear{Mayor \& Queloz}{Mayor \&
  Queloz}{1995}]{mayor1995jupiter}
Mayor M.,  Queloz D.,  1995, A Jupiter-mass companion to a solar-type star

\bibitem[\protect\citeauthoryear{{Owen} \& {Jackson}}{{Owen} \&
  {Jackson}}{2012}]{owen_2012}
{Owen} J.~E.,  {Jackson} A.~P.,  2012, \mn@doi [\mnras]
  {10.1111/j.1365-2966.2012.21481.x}, \href
  {http://adsabs.harvard.edu/abs/2012MNRAS.425.2931O} {425, 2931}

\bibitem[\protect\citeauthoryear{Poppenhaeger, Schmitt  \& Wolk}{Poppenhaeger
  et~al.}{2013}]{poppenhaeger2013transit}
Poppenhaeger K.,  Schmitt J.,   Wolk S.,  2013, The Astrophysical Journal, 773,
  62

\bibitem[\protect\citeauthoryear{Ribas, Guinan, G{\"u}del  \& Audard}{Ribas
  et~al.}{2005}]{ribas2005evolution}
Ribas I.,  Guinan E.~F.,  G{\"u}del M.,   Audard M.,  2005, The Astrophysical
  Journal, 622, 680

\bibitem[\protect\citeauthoryear{Salz, Schneider, Czesla  \& Schmitt}{Salz
  et~al.}{2016a}]{salz2016energy}
Salz M.,  Schneider P.,  Czesla S.,   Schmitt J.,  2016a, Astronomy \&
  Astrophysics, 585, L2

\bibitem[\protect\citeauthoryear{{Salz}, {Czesla}, {Schneider}  \&
  {Schmitt}}{{Salz} et~al.}{2016b}]{salz_2016a}
{Salz} M.,  {Czesla} S.,  {Schneider} P.~C.,   {Schmitt} J.~H.~M.~M.,  2016b,
  \mn@doi [\aap] {10.1051/0004-6361/201526109}, \href
  {http://adsabs.harvard.edu/abs/2016A%26A...586A..75S} {586, A75}

\bibitem[\protect\citeauthoryear{{Sanchis-Ojeda} \& {Winn}}{{Sanchis-Ojeda} \&
  {Winn}}{2011}]{ojeda_2011}
{Sanchis-Ojeda} R.,  {Winn} J.~N.,  2011, \mn@doi [\apj]
  {10.1088/0004-637X/743/1/61}, \href
  {http://adsabs.harvard.edu/abs/2011ApJ...743...61S} {743, 61}

\bibitem[\protect\citeauthoryear{{Sanchis-Ojeda} et~al.,}{{Sanchis-Ojeda}
  et~al.}{2013}]{ojeda_2013}
{Sanchis-Ojeda} R.,  et~al., 2013, \mn@doi [\apj] {10.1088/0004-637X/775/1/54},
  \href {http://adsabs.harvard.edu/abs/2013ApJ...775...54S} {775, 54}

\bibitem[\protect\citeauthoryear{Sanz-Forcada, Ribas, Micela, Pollock,
  Garc{\'\i}a-{\'A}lvarez, Solano  \& Eiroa}{Sanz-Forcada
  et~al.}{2010}]{sanz2010scenario}
Sanz-Forcada J.,  Ribas I.,  Micela G.,  Pollock A.,  Garc{\'\i}a-{\'A}lvarez
  D.,  Solano E.,   Eiroa C.,  2010, Astronomy \& Astrophysics, 511, L8

\bibitem[\protect\citeauthoryear{Sanz-Forcada, Micela, Ribas, Pollock, Eiroa,
  Velasco, Solano  \& Garc{\'\i}a-{\'A}lvarez}{Sanz-Forcada
  et~al.}{2011}]{sanz2011estimation}
Sanz-Forcada J.,  Micela G.,  Ribas I.,  Pollock A.~M.,  Eiroa C.,  Velasco A.,
   Solano E.,   Garc{\'\i}a-{\'A}lvarez D.,  2011, Astronomy \& Astrophysics,
  532, A6

\bibitem[\protect\citeauthoryear{{Sanz-Forcada}, {Desidera}  \&
  {Micela}}{{Sanz-Forcada} et~al.}{2014}]{sanz_2014}
{Sanz-Forcada} J.,  {Desidera} S.,   {Micela} G.,  2014, \mn@doi [\aap]
  {10.1051/0004-6361/201323231}, \href
  {http://adsabs.harvard.edu/abs/2014A%26A...570A..50S} {570, A50}

\bibitem[\protect\citeauthoryear{{Schr{\"o}ter}, {Czesla}, {Wolter},
  {M{\"u}ller}, {Huber}  \& {Schmitt}}{{Schr{\"o}ter}
  et~al.}{2011}]{schroeter_2011}
{Schr{\"o}ter} S.,  {Czesla} S.,  {Wolter} U.,  {M{\"u}ller} H.~M.,  {Huber}
  K.~F.,   {Schmitt} J.~H.~M.~M.,  2011, \mn@doi [\aap]
  {10.1051/0004-6361/201116961}, \href
  {http://adsabs.harvard.edu/abs/2011A%26A...532A...3S} {532, A3}

\bibitem[\protect\citeauthoryear{{Shematovich}, {Ionov}  \&
  {Lammer}}{{Shematovich} et~al.}{2014}]{shematovich_2014}
{Shematovich} V.~I.,  {Ionov} D.~E.,   {Lammer} H.,  2014, \mn@doi [\aap]
  {10.1051/0004-6361/201423573}, \href
  {http://adsabs.harvard.edu/abs/2014A%26A...571A..94S} {571, A94}

\bibitem[\protect\citeauthoryear{{Smith}, {Brickhouse}, {Liedahl}  \&
  {Raymond}}{{Smith} et~al.}{2001}]{smith_2001}
{Smith} R.~K.,  {Brickhouse} N.~S.,  {Liedahl} D.~A.,   {Raymond} J.~C.,  2001,
  in {Ferland} G.,  {Savin} D.~W.,  eds,  Astronomical Society of the Pacific
  Conference Series Vol. 247, Spectroscopic Challenges of Photoionized Plasmas.
  p.~161

\bibitem[\protect\citeauthoryear{{Spiegel}, {Fortney}  \& {Sotin}}{{Spiegel}
  et~al.}{2014}]{spiegel_2014}
{Spiegel} D.~S.,  {Fortney} J.~J.,   {Sotin} C.,  2014, \mn@doi [Proceedings of
  the National Academy of Science] {10.1073/pnas.1304206111}, \href
  {http://adsabs.harvard.edu/abs/2014PNAS..11112622S} {111, 12622}

\bibitem[\protect\citeauthoryear{{Swift} et~al.,}{{Swift}
  et~al.}{2012}]{swift_2012}
{Swift} D.~C.,  et~al., 2012, \mn@doi [\apj] {10.1088/0004-637X/744/1/59},
  \href {http://adsabs.harvard.edu/abs/2012ApJ...744...59S} {744, 59}

\bibitem[\protect\citeauthoryear{Valencia, Ikoma, Guillot  \&
  Nettelmann}{Valencia et~al.}{2010}]{valencia2010composition}
Valencia D.,  Ikoma M.,  Guillot T.,   Nettelmann N.,  2010, Astronomy \&
  Astrophysics, 516, A20

\bibitem[\protect\citeauthoryear{Vidal-Madjar, Des~Etangs, D{\'e}sert,
  Ballester, Ferlet, H{\'e}brard  \& Mayor}{Vidal-Madjar
  et~al.}{2003}]{vidal2003extended}
Vidal-Madjar A.,  Des~Etangs A.~L.,  D{\'e}sert J.-M.,  Ballester G.,  Ferlet
  R.,  H{\'e}brard G.,   Mayor M.,  2003, Nature, 422, 143

\bibitem[\protect\citeauthoryear{Watson, Donahue  \& Walker}{Watson
  et~al.}{1981}]{watson1981dynamics}
Watson A.~J.,  Donahue T.~M.,   Walker J.~C.,  1981, Icarus, 48, 150

\bibitem[\protect\citeauthoryear{{Welsh}, {Lallement}, {Vergely}  \&
  {Raimond}}{{Welsh} et~al.}{2010}]{welsh_2010}
{Welsh} B.~Y.,  {Lallement} R.,  {Vergely} J.-L.,   {Raimond} S.,  2010,
  \mn@doi [\aap] {10.1051/0004-6361/200913202}, \href
  {http://adsabs.harvard.edu/abs/2010A%26A...510A..54W} {510, A54}

\bibitem[\protect\citeauthoryear{Worthey \& Lee}{Worthey \&
  Lee}{2011}]{worthey2011empirical}
Worthey G.,  Lee H.-c.,  2011, The Astrophysical Journal Supplement Series,
  193, 1

\makeatother
\end{thebibliography}
\end{document}